\documentclass[aps,pra,superscriptaddress,amsmath,amssymb,preprintnumbers,showpacs,twocolumn]{revtex4}
\usepackage{amssymb}
\usepackage{graphicx}
\usepackage{subfigure}
\usepackage{bm}
\usepackage{color}

\newcommand{\Ignore}[1]{}

\newcommand{\Ket}[1]{\left\vert #1\right\rangle}

\newcommand{\KetBra}[2]{\left\vert#1\right\rangle\left\langle#2\right\vert}
\newcommand{\Projector}[1]{\KetBra{#1}{#1}}

\renewcommand{\eqref}[1]{(\ref{#1})} 

\begin{document}
\title{Stimulated Raman adiabatic passage in a $\Lambda$-system in the presence of quantum noise}

\author{M. Scala}
\affiliation{Dipartimento di Scienze Fisiche ed Astronomiche dell'Universit\`{a} di Palermo, Via Archirafi 36, 90123 Palermo, Italy}

\author{B. Militello}
\affiliation{Dipartimento di Scienze Fisiche ed Astronomiche dell'Universit\`{a} di Palermo, Via Archirafi 36, 90123 Palermo, Italy}

\author{A. Messina}
\affiliation{Dipartimento di Scienze Fisiche ed Astronomiche dell'Universit\`{a} di Palermo, Via Archirafi 36, 90123 Palermo, Italy}

\author{N. V. Vitanov}
\affiliation{Department of Physics, Sofia University, James Bourchier 5 blvd, 1164 Sofia, Bulgaria}

\begin{abstract}
We exploit a microscopically derived master equation for the study
of STIRAP in the presence of decay from the auxiliary level toward
the initial and final state, and compare our results with the
predictions obtained from a phenomenological model previously used
[P. A. Ivanov, N. V. Vitanov, and K. Bergmann, Phys. Rev. A {\bf
72}, 053412 (2005)]. It is shown that our approach predicts a much
higher efficiency. The effects of temperature are also taken into
account, proving that in b-STIRAP thermal pumping can increase the
efficiency of the population transfer.
\end{abstract}


\maketitle


\section{Introduction}

Adiabatic theorem \cite{ref:Messiah} provides a very powerful tool
for quantum state manipulation, and indeed it has been the basis
of many applications, aimed at generation of quantum states
\cite{ref:Cirac} or at the realization of quantum gates
\cite{ref:GeomPhases-exp} based on geometric phase
\cite{ref:GeomPhases}. Even if the wide range of validity of the
adiabatic theorem has been recently criticized
\cite{ref:Marzlin2004}, very recently a sufficient condition that
guarantees adiabatic evolutions has been proven
\cite{ref:Comparat2009}.

Based on the adiabatic theorem, stimulated Raman adiabatic passage
(STIRAP) \cite{ref:STIRAP_original_1, ref:STIRAP_original_2}
allows the transfer of population from a quantum state of a
physical system toward another state, through an auxiliary
intermediate state \cite{ref:STIRAP_reviews}. The passage occurs
via a {\it dark} state, which aligns with the initial state in the
beginning of the process, and then gradually changes its structure
toward alignment with the target state. The process ends when the
dark and target states coincide. For the opposite delay of the two
couplings between the three states, and for nonzero single-photon
detuning, it is also possible to realize the b-STIRAP process,
which instead exploits the adiabatic change of a {\it bright}
eigenstate of the Hamiltonian from the initial state to the target
state. The main difference between STIRAP and b-STIRAP is that in
the latter case the auxiliary state is effectively involved in the
dynamics, in the sense that a certain amount of population is
temporarily transferred to it during the process. This
circumstance makes the b-STIRAP more sensitive than STIRAP to the
presence of decay from the auxiliary level. In fact, while in the
absence of environmental interaction and classical noise the
transfer from the initial state to the target state is predicted
to be perfect, in the presence of dissipative dynamics the
efficiency of the process is negatively affected. Many different
models have been considered to study the effect of dephasing
\cite{STIRAP dephasing} and spontaneous emission from the
auxiliary level either toward external states
\cite{ref:Vitanov1997} or toward internal states \cite{STIRAP
spontaneous emission}, i.e. the initial and the target state. In
all these models, the incoherent dynamics has been taken into
account phenomenologically. Very recently, a microscopic model to
describe the STIRAP and b-STIRAP processes under dissipation from
the auxiliary state toward external states has been presented
\cite{ref:Scala2010}. The derivation of the master equation from a
model of interaction between the three-state system and a bosonic
environment, and the relevant dynamics, show some interesting
deviations from the predictions related to the phenomenological
counterpart. In particular, the microscopic model predicts a much
higher efficiency in the STIRAP scheme.

In this paper, by using the same rigorous microscopic model as
described above \cite{ref:Scala2010},
 we investigate the effect of spontaneous decay from the intermediate state inside the $\Lambda$-system.
Because the initial and final states in STIRAP are usually ground
or metastable, the intermediate state is necessarily an excited
state, which may decay both inside and outside the system
\cite{Pillet93,Goldner94,Chu94,Halfmann96,Theuer98}. The loss of
efficiency caused by external decay is more detrimental for it
leads to irreversible population loss from the system; it is also
easier to describe and understand
\cite{ref:Vitanov1997,ref:Scala2010}. The effect of internal decay
on STIRAP is a much more subtle effect because the loss of
efficiency is compensated by the concomitant optical pumping
\cite{STIRAP spontaneous emission}. Here we develop a rigorous
microscopic theory of internal decay in STIRAP and b-STIRAP, which
reveals some unexpected features compared to the phenomenological
model \cite{STIRAP spontaneous emission}.

Starting from a microscopic model of system-environment
interaction, we derive a time-dependent master equation that
describes the dynamics of our system. Then, after considering the
resolution of the master equation, we compare the predictions
coming from our model with the results coming from the
phenomenological description of an analogous decay scheme
\cite{STIRAP spontaneous emission}. We find that the efficiency of
the STIRAP process is higher than predicted before. Exploiting the
master equation at nonzero temperature, we also study the effects
of temperature, showing that the thermal pumping dramatically and
negatively affects the efficiency of the population transfer in
the STIRAP process, while has a slightly positive effect in
b-STIRAP.

The paper is organized as follows. In the next section we present
the derivation of the Markovian master equation of a system with
time-independent Hamiltonian and a time-dependent
system-environment interaction term. In the third section we apply
the result of the previous section and the general theory of
Davies and Spohn \cite{ref:Davies1978,ref:Florio2006} to derive
the master equation of out three-state system. Then, in the fourth
section we show the results obtained at zero temperature and
compare them with the results coming from the phenomenological
model. In section V we show the effects of temperature, and
finally in the last section we give some conclusive remarks.

\section{General formalism}

In this section we consider the general problem of the derivation
of the master equation for a system whose Hamiltonian $H_s$ is
constant, while the system-environment interaction Hamiltonian
contains oscillation terms. This will be useful in the next
section, where we deal with a system described (in a rotating
frame) by a slowly-varying Hamiltonian and interacting with a
thermal bath through oscillating terms. According to the general
theory by Davies and Spohn \cite{ref:Davies1978,ref:Florio2006},
under the hypothesis that the environmental correlation time is
much smaller than the timescale of the Hamiltonian change, we can
treat this system by assuming that the system Hamiltonian is
time-independent during the derivation of the master equation, and
putting the time-dependence of the jump operators only after the
derivation.

Therefore we start by considering a time-independent system
Hamiltonian $H_s$ and the following time-dependent system-bath
interaction Hamiltonian:
\begin{equation}
  H_I = \sum_\alpha \left(A_\alpha^+\, e^{i\omega_\alpha\,t} + A_\alpha^-\, e^{-i\omega_\alpha\,t}\right) \otimes
  B_\alpha\,.
\end{equation}

Following the approach presented in \cite{ref:Petru}, let us
introduce, for each Bohr frequency $\omega$
\begin{equation}
  A_\alpha^\pm(\omega) = \sum_{\epsilon'-\epsilon=\omega}
  \Pi(\epsilon)A_\alpha^\pm\Pi(\epsilon')\,,
\end{equation}
where $\Pi(\epsilon)$ is the projector on the subspace of the
system Hilbert space corresponding to the energy eigenvalue
$\epsilon$ and the sum is extended over all the couples of
energies $\epsilon$ and $\epsilon'$ such that
$\epsilon'-\epsilon=\omega$. The operators defined in this way
satisfy both
\begin{equation}
  [H_s,A_\alpha^\pm(\omega)] = -\omega_\alpha A_\alpha^\pm(\omega)\,,
\end{equation}
and
\begin{equation}
  \left(A_\alpha^\pm(\omega)\right)^\dag = A_\alpha^\mp(-\omega)\,,
\end{equation}
giving
\begin{equation}
  e^{i\,H_s\,t}\,A_\alpha^\pm(\omega)\,e^{-i\,H_s\,t} = e^{-i\,\omega\,t}\,A_\alpha^\pm(\omega),
\end{equation}
\begin{equation}
  e^{i\,H_s\,t}\,\left(A_\alpha^\pm(\omega)\right)^\dag\,e^{-i\,H_s\,t} =
  e^{i\,\omega\,t}\,\left(A_\alpha^\pm(\omega)\right)^\dag\,.
\end{equation}

Another important property is that summing over all the Bohr
frequencies (both positive and negative) one reobtains the initial
operators:
\begin{equation}
A_\alpha^{\pm} = \sum_\omega\, A_\alpha^{\pm}(\omega)\,.
\end{equation}

In the Schr\"odinger picture we thus have:
\begin{equation} H_I =
\sum_{\alpha,\omega}
\left(A_\alpha^+(\omega)\,e^{i\omega_\alpha\,t}+A_\alpha^-(\omega)\,e^{-i\omega_\alpha\,t}\right)\otimes
B_\alpha\,,
\end{equation}
which in the interaction picture with respect to $H_s+H_B$
becomes:
\begin{equation} H_I =
\sum_{\alpha,\omega}
e^{-i\,\omega\,t}\,\left(A_\alpha^+(\omega)\,e^{i\omega_\alpha\,t}+A_\alpha^-(\omega)\,e^{-i\omega_\alpha\,t}\right)\otimes
B_\alpha(t)\,,
\end{equation}
or, taking the Hermitian conjugate:
\begin{eqnarray}
\nonumber H_I &=& \sum_{\alpha,\omega}
e^{i\,\omega\,t}\,\left(\left(A_\alpha^+(\omega)\right)^\dag\,e^{-i\omega_\alpha\,t}+\left(A_\alpha^-(\omega)\right)^\dag\,e^{i\omega_\alpha\,t}\right)\\
&\otimes &
B_\alpha^\dag(t)\,.
\end{eqnarray}

The formal resolution of the Liouville equation gives:
\begin{widetext}
\begin{equation}
\dot\rho = \int_0^\infty \mathrm{d}s\,\mathrm{tr}_B
\left\{H_I(t-s)\rho(t)\rho_B H_I(t)-H_I(t)H_I(t-s)\rho(t)\rho_B
\right\} + h.c.\,,
\end{equation}
\end{widetext}
from which, substituting the expansions of $H_I$, one gets the
following master equation:
\begin{widetext}
\begin{eqnarray} \nonumber \dot\rho &=&
\sum_{\omega,\omega'}\sum_{\alpha,\beta}
e^{i(\omega-\omega'+\omega_\beta-\omega_\alpha)t}\,\Gamma_{\alpha\beta}^{++}(\omega)\left(
A_\beta^+(\omega)\rho \left(A_\alpha^+(\omega')\right)^\dag -
\left(A_\alpha^+(\omega')\right)^\dag A_\beta^+(\omega)\rho
\right)\\
\nonumber
&+& \sum_{\omega,\omega'}\sum_{\alpha,\beta}
e^{i(\omega-\omega'+\omega_\beta+\omega_\alpha)t}\,\Gamma_{\alpha\beta}^{-+}(\omega)\left(
A_\beta^+(\omega)\rho \left(A_\alpha^-(\omega')\right)^\dag -
\left(A_\alpha^-(\omega')\right)^\dag A_\beta^+(\omega)\rho
\right)\\
\nonumber &+& \sum_{\omega,\omega'}\sum_{\alpha,\beta}
e^{i(\omega-\omega'-\omega_\beta-\omega_\alpha)t}\,\Gamma_{\alpha\beta}^{+-}(\omega)\left(
A_\beta^-(\omega)\rho \left(A_\alpha^+(\omega')\right)^\dag -
\left(A_\alpha^+(\omega')\right)^\dag A_\beta^-(\omega)\rho
\right)\\
&+& \sum_{\omega,\omega'}\sum_{\alpha,\beta}
e^{i(\omega-\omega'-\omega_\beta+\omega_\alpha)t}\,\Gamma_{\alpha\beta}^{--}(\omega)\left(
A_\beta^-(\omega)\rho \left(A_\alpha^-(\omega')\right)^\dag -
\left(A_\alpha^-(\omega')\right)^\dag A_\beta^-(\omega)\rho
\right) + h.c.\,,
\end{eqnarray}
\end{widetext} with
\begin{eqnarray}\label{eq:Gamma++}
\nonumber
\Gamma_{\alpha\beta}^{++}(\omega)&=&\Gamma_{\alpha\beta}^{-+}(\omega)\\
&=&\int_0^\infty \mathrm{d}s\, e^{i(\omega-\omega_\beta)\,s}
\left\langle B_\alpha^\dag(t)\,B_\beta(t-s)\right\rangle\,,
\end{eqnarray}
and
\begin{eqnarray}\label{eq:Gamma--}
\nonumber
\Gamma_{\alpha\beta}^{+-}(\omega)&=&\Gamma_{\alpha\beta}^{--}(\omega)\\
&=&\int_0^\infty \mathrm{d}s\, e^{i(\omega+\omega_\beta)\,s}
\left\langle B_\alpha^\dag(t)\,B_\beta(t-s)\right\rangle\,.
\end{eqnarray}

This is the most general form of the Born-Markov master equation
before a Rotating Wave Approximation (RWA) is performed. Under the
hypothesis that $\omega_\alpha,\omega_\beta \gg \omega,\omega'$,
one can single out very clear conditions for RWA. The only terms
which survive are those for which $\omega_\alpha$ and
$\omega_\beta$ appear in the combination
$\omega_\alpha-\omega_\beta$ with $\alpha=\beta$, and
$\omega=\omega'$:
\begin{eqnarray}
\nonumber \dot\rho &=& \sum_{\omega}\sum_{\alpha}
\,\Gamma_{\alpha\alpha}^{++}(\omega)\left( A_\alpha^+(\omega)\rho
\left(A_\alpha^+(\omega)\right)^\dag\right.\\
\nonumber &-& \left.\left(A_\alpha^+(\omega)\right)^\dag
A_\alpha^+(\omega)\rho
\right)\\
\nonumber &+& \sum_{\omega}\sum_{\alpha}
\,\Gamma_{\alpha\alpha}^{--}(\omega)\left( A_\alpha^-(\omega)\rho
\left(A_\alpha^-(\omega)\right)^\dag\right. \\
&-& \left.\left(A_\alpha^-(\omega)\right)^\dag
A_\alpha^-(\omega)\rho \right) + h.c.\,
\end{eqnarray}
which, neglecting the Lamb shifts and coming back to the
Schr\"odinger picture, becomes:
\begin{eqnarray}
\label{eq:ME_General_Lindb} \nonumber \dot\rho &=&
-i[H_s,\rho]+\sum_{\omega}\sum_{\alpha}
\,\gamma_{\alpha\alpha}^{++}(\omega) \left( A_\alpha^+(\omega)\rho
\left(A_\alpha^+(\omega)\right)^\dag \right.\\
\nonumber
&-& \left.
\frac{1}{2}\left\{\left(A_\alpha^+(\omega)\right)^\dag
A_\alpha^+(\omega),\rho\right\}
\right)\\
\nonumber
&+& \sum_{\omega}\sum_{\alpha}
\,\gamma_{\alpha\alpha}^{--}(\omega) \left( A_\alpha^-(\omega)\rho
\left(A_\alpha^-(\omega)\right)^\dag\right.\\
&-& \left. \frac{1}{2}\left\{\left(A_\alpha^-(\omega)\right)^\dag
A_\alpha^-(\omega),\rho\right\} \right)\,,
\end{eqnarray}
where
$\gamma_{\alpha\alpha}^{++}(\omega)=2\Re\{\Gamma_{\alpha\alpha}^{++}(\omega)\}$
and
$\gamma_{\alpha\alpha}^{--}(\omega)=2\Re\{\Gamma_{\alpha\alpha}^{--}(\omega)\}$.

\section{Our model}

\subsection{The system}

We consider a three-level system in $\Lambda$-configuration whose
Hamiltonian is:
\begin{widetext}
\begin{equation}\label{eq:Hsys}
H_{sys}(t) = \left[\begin{array}{ccc}
\omega_1 & \Omega_p(t) e^{i(\omega_{21}-\Delta)t} & 0  \\
\Omega_p(t)e^{-i(\omega_{21}-\Delta)t} & \omega_2 & \Omega_s(t)e^{-i(\omega_{23}-\Delta)t} \\
0 & \Omega_s(t)e^{i(\omega_{23}-\Delta)t} & \omega_3  \\
\end{array}\right]\,.
\end{equation}
\end{widetext}

\begin{figure}
\includegraphics[width=0.40\textwidth, angle=0]{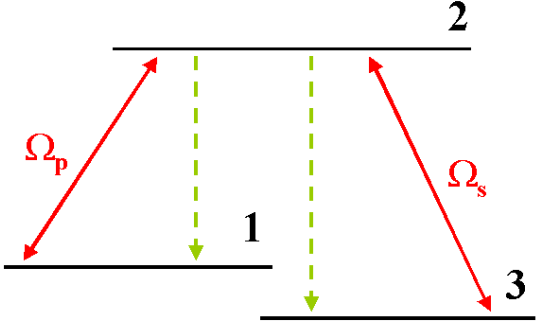}
\caption{(Color online). States $\Ket{1}$ and $\Ket{3}$ are
coherently coupled to level $\Ket{2}$ (red bold lines). The state
$\Ket{2}$ is coupled to the other two states by a dipolar
system-environment interaction (green dashed lines).}
\label{fig:system}
\end{figure}

The system interacts with a bosonic bath. The free bath is
described by
\begin{equation}
H_B = \sum_k\omega_k\, b_k^\dag b_k\,,
\end{equation}
while the system-bath interaction Hamiltonian is
\begin{eqnarray}\label{Hinteraction}
\nonumber H_{int} &=& \sum_k g_k^{(12)}
\left(\KetBra{1}{2}+\KetBra{2}{1}\right)(b_k+b_k^\dag)\\
&+& \sum_k g_k^{(32)}
\left(\KetBra{3}{2}+\KetBra{2}{3}\right)(b_k+b_k^\dag)\,.
\end{eqnarray}

In the rotating frame associated with the transformation
$T(t)=e^{i\omega_{1}t}\Projector{1}+e^{i(\omega_{2}-\Delta)t}\Projector{2}+e^{i\omega_{3}t}\Projector{3}$,
the total Hamiltonian:
\begin{equation}
H = H_s(t)+H_B+H_I(t)\,,
\end{equation}
with
\begin{equation}\label{eq:HSystem} H_s(t) =
\left[\begin{array}{ccc}
0 & \Omega_p(t) & 0  \\
\Omega_p(t) & \Delta & \Omega_s(t) \\
0 & \Omega_s(t) & 0  \\
\end{array}\right]\,,
\end{equation}
and
\begin{eqnarray}\label{eq:HInteraction}
\nonumber H_I(t) &=& \sum_k g_k^{(12)}
\left(e^{i(\omega_1-\omega_2+\Delta)\,t}\,\KetBra{1}{2}\right.\\
\nonumber &+& \left. e^{-i(\omega_1-\omega_2+\Delta)\,t}\,\KetBra{2}{1}\right)(b_k+b_k^\dag)\\
\nonumber &+& \sum_k g_k^{(32)}
\left(e^{i(\omega_3-\omega_2+\Delta)\,t}\,\KetBra{3}{2}\right. \\
&+& \left.
e^{-i(\omega_3-\omega_2+\Delta)\,t}\,\KetBra{2}{3}\right)(b_k+b_k^\dag)\,.
\end{eqnarray}

The eigenstates and eigenvalues of $H_s(t)$ are:
\begin{subequations}
\begin{equation}
\omega_+=\Omega_0 \cot\varphi,\,\,\,\, \omega_0=0,\,\,\,\,
\omega_-=-\Omega_0\tan\varphi,
\end{equation}
\begin{align}
\Ket{+}&= \sin\varphi\sin\theta\Ket{1}+\cos\varphi\Ket{2}+\sin\varphi\cos\theta\Ket{3}\,,\\
\Ket{0}&= \cos\theta\Ket{1}-\sin\theta\Ket{3}\,,\\
\Ket{-}&=
\cos\varphi\sin\theta\Ket{1}-\sin\varphi\Ket{2}+\cos\varphi\cos\theta\Ket{3}\,,
\end{align}
\end{subequations}
where
\begin{subequations}
\begin{align}
\tan\theta(t)&=\frac{\Omega_p(t)}{\Omega_s(t)}\,,\\
\tan{2\varphi(t)}&=\frac{2\Omega(t)}{\Delta(t)}\,,\\
\Omega(t)&=\sqrt{\Omega_p(t)^2+\Omega_s(t)^2}\,.
\end{align}
\end{subequations}

It is well known \cite{ref:STIRAP_reviews} that, for the intuitive
sequence of pulses, in which the probe pulse $\Omega_p(t)$
precedes the Stokes pulse $\Omega_s(t)$, one has $\Ket{-}=\Ket{1}$
for $t\rightarrow -\infty$ and $\Ket{-}=\Ket{3}$ for $t\rightarrow
\infty$. Therefore, if all the pulses vary adiabatically, the
population of the state $\Ket{1}$ can be transferred to the state
$\Ket{3}$: this process is called b-STIRAP. On the other hand, if
$\Omega_s(t)$ precedes $\Omega_p(t)$, one has $\Ket{0}=\Ket{1}$
for $t\rightarrow -\infty$ and $\Ket{0}=\Ket{3}$ for $t\rightarrow
\infty$: in this counterintuitive sequence the population from
$\Ket{1}$ to $\Ket{3}$ is adiabatically transferred through the
state $\Ket{0}$ and the process is called STIRAP.

\subsection{Master Equation}
Since in the rotating frame we have a slowly-varying system
Hamiltonian and a time-dependent system-environment interaction
term, we can use the general formalism presented before in order
to derive the master equation that describes the dynamics of our
system. We obtain the following master equation (for details see
appendix \ref{AppDerivation}):
\begin{widetext}
\begin{eqnarray}\label{eq:MasterEq_NonZeroTemperature}
\nonumber \dot\rho &=& -i[H_s,\rho]+
\,\left[\gamma_{aa}^{++}(\omega_{+0})\cos^2\theta\cos^2\varphi+
\gamma_{bb}^{++}(\omega_{+0})\sin^2\theta\cos^2\varphi\right]
\left(\KetBra{0}{+}\rho \KetBra{+}{0} -
\frac{1}{2}\left\{\KetBra{+}{+},\rho\right\}\right)\\
\nonumber &+&
\,\left[\gamma_{aa}^{--}(\omega_{0-})\cos^2\theta\sin^2\varphi+
\gamma_{bb}^{--}(\omega_{0-})\sin^2\theta\sin^2\varphi\right]
\left(\KetBra{-}{0}\rho \KetBra{0}{-} -
\frac{1}{2}\left\{\KetBra{0}{0},\rho\right\}\right)\\
\nonumber &+&
\,\left[\gamma_{aa}^{++}(\omega_{+-})\sin^2\theta\cos^4\varphi
+\gamma_{aa}^{--}(\omega_{+-})\sin^2\theta\sin^4\varphi
+\gamma_{bb}^{++}(\omega_{+-})\cos^2\theta\cos^4\varphi
+\gamma_{bb}^{--}(\omega_{+-})\cos^2\theta\sin^4\varphi\right]\\
\nonumber &&\times \left(\KetBra{-}{+}\rho \KetBra{+}{-} -
\frac{1}{2}\left\{\KetBra{+}{+},\rho\right\}\right)\\
\nonumber &+&
\,\left[\left(\gamma_{aa}^{++}(0)+\gamma_{aa}^{--}(0)\right)\sin^2\theta\sin^2\varphi\cos^2\varphi+
\left(\gamma_{bb}^{++}(0)+\gamma_{bb}^{--}(0)\right)\cos^2\theta\sin^2\varphi\cos^2\varphi\right]\nonumber\\
\nonumber &&\times
\left[\left(\KetBra{+}{+}-\KetBra{-}{-}\right)\rho\left(\KetBra{+}{+}-\KetBra{-}{-}\right)
-
\frac{1}{2}\left\{\KetBra{+}{+}+\KetBra{-}{-},\rho\right\}\right]\\
\nonumber &+&
\,\left[\gamma_{aa}^{--}(\omega_{0+})\cos^2\theta\cos^2\varphi+
\gamma_{bb}^{--}(\omega_{0+})\sin^2\theta\cos^2\varphi\right]
\left(\KetBra{+}{0}\rho \KetBra{0}{+} -
\frac{1}{2}\left\{\KetBra{0}{0},\rho\right\}\right)\\
\nonumber &+&
\,\left[\gamma_{aa}^{++}(\omega_{-0})\cos^2\theta\sin^2\varphi+
\gamma_{bb}^{++}(\omega_{-0})\sin^2\theta\sin^2\varphi\right]
\left(\KetBra{0}{-}\rho \KetBra{-}{0} -
\frac{1}{2}\left\{\KetBra{-}{-},\rho\right\}\right)\\
\nonumber &+&
\,\left[\gamma_{aa}^{--}(\omega_{-+})\sin^2\theta\cos^4\varphi+
\gamma_{aa}^{++}(\omega_{-+})\sin^2\theta\sin^4\varphi+
\gamma_{bb}^{--}(\omega_{-+})\cos^2\theta\cos^4\varphi+
\gamma_{bb}^{++}(\omega_{-+})\cos^2\theta\sin^4\varphi\right]\\
&&\times \left(\KetBra{+}{-}\rho \KetBra{-}{+} -
\frac{1}{2}\left\{\KetBra{-}{-},\rho\right\}\right)\,,
\end{eqnarray}
\end{widetext}
where $\omega_{nm}=\omega_n-\omega_m$.

From \eqref{eq:Gamma++} and \eqref{eq:Gamma--} one gets that the
decay rates are given by a spectral density $J_j(\omega)$
multiplied by a factor depending on the photon population
$N(\omega)$ of the bath modes at the relevant frequency corrected
with $\pm\omega_a$ or $\pm\omega_b$ depending on the case, i.e.:
\begin{eqnarray}\label{eq:EnneDiOmega}
\nonumber \!\! \left\{
\begin{array}{ll}
\gamma_{jj}^{++}(\omega) = J_j(\omega-\omega_j) (1+N(\omega-\omega_j)) \qquad &\omega-\omega_j>0\,, \\
\gamma_{jj}^{--}(\omega) = J_j(\omega+\omega_j)
(1+N(\omega+\omega_j)) \qquad &\omega+\omega_j>0 \,, \\
\gamma_{jj}^{++}(\omega) = J_j(|\omega-\omega_j|)\,
N(|\omega-\omega_j|)  \qquad &\omega-\omega_j<0 \,,\\
\gamma_{jj}^{--}(\omega) = J_j(|\omega+\omega_j|)\,
N(|\omega+\omega_j|)  \qquad &\omega+\omega_j<0 \,,
\end{array}
\right.\\
\end{eqnarray}
with $j=a,b$ and
\begin{subequations}
\begin{eqnarray}\label{eq:frequencies}
\omega_{a}=\omega_1-\omega_2+\Delta\,,\\
\omega_{b}=\omega_3-\omega_2+\Delta\,.
\end{eqnarray}
\end{subequations}

The zero temperature spectral density $J_j(\omega)$ for general
bosonic reservoir is given by \cite{ref:Gardiner, ref:Petru}:
\begin{eqnarray}\label{Jomega}
 J_j(\omega)=d(\omega)\,\left|g_j(\omega)\right|^2
\end{eqnarray}
where $g_a(\omega)$ ($g_b(\omega)$) is the system-reservoir
coupling constant $g_k^{(12)}$ ($g_k^{(32)}$) in the continuum
limit, and $d(\omega)$ is the reservoir density of states at
frequency $\omega$.

It is important to note that, under the hypothesis that
$\omega_j\gg \omega$ for any Bohr frequency $\omega$ between the
dressed states in the rotating frame (which is the usual case
since $\omega_j$ are optical frequencies associated with the
atomic transitions while $\omega$'s are of the order of magnitude
of the coupling terms $\Omega$'s), and taking into account that
the frequencies in \eqref{eq:frequencies} are negative, the only
condition satisfied are $\omega-\omega_j>0$ and
$\omega+\omega_j<0$. Therefore, in (\ref{eq:EnneDiOmega}), only
the rates of the first and fourth classes are possible. Moreover,
at zero temperature only the rates of the first class survive,
since the number of photons in the reservoir is zero. In such a
case the master equation becomes:
\begin{widetext}
\begin{eqnarray}\label{eq:MasterEq_ZeroTemperature}
\nonumber \dot\rho &=& -i[H_s,\rho]+
\,\left[\gamma_{aa}^{++}(\omega_{+0})\cos^2\theta\cos^2\varphi+
\gamma_{bb}^{++}(\omega_{+0})\sin^2\theta\cos^2\varphi\right]
\left(\KetBra{0}{+}\rho \KetBra{+}{0} -
\frac{1}{2}\left\{\KetBra{+}{+},\rho\right\}\right)\\
\nonumber &+&
\,\left[\gamma_{aa}^{++}(\omega_{-0})\cos^2\theta\sin^2\varphi+
\gamma_{bb}^{++}(\omega_{-0})\sin^2\theta\sin^2\varphi\right]
\left(\KetBra{0}{-}\rho \KetBra{-}{0} -
\frac{1}{2}\left\{\KetBra{-}{-},\rho\right\}\right)\\
\nonumber &+&
\,\left[\gamma_{aa}^{++}(\omega_{+-})\sin^2\theta\cos^4\varphi
+\gamma_{bb}^{++}(\omega_{+-})\cos^2\theta\cos^4\varphi\right]
\nonumber \times \left(\KetBra{-}{+}\rho \KetBra{+}{-} -
\frac{1}{2}\left\{\KetBra{+}{+},\rho\right\}\right)\\
\nonumber &+& \,\left[
\gamma_{aa}^{++}(\omega_{-+})\sin^2\theta\sin^4\varphi+
\gamma_{bb}^{++}(\omega_{-+})\cos^2\theta\sin^4\varphi\right]
\nonumber \times \left(\KetBra{+}{-}\rho \KetBra{-}{+} -
\frac{1}{2}\left\{\KetBra{-}{-},\rho\right\}\right)\\
&+&
\,\left[\gamma_{aa}^{++}(0)\sin^2\theta\sin^2\varphi\cos^2\varphi+
\gamma_{bb}^{++}(0)\cos^2\theta\sin^2\varphi\cos^2\varphi\right]\nonumber\\
&&\times
\left[\left(\KetBra{+}{+}-\KetBra{-}{-}\right)\rho\left(\KetBra{+}{+}-\KetBra{-}{-}\right)
-
\frac{1}{2}\left\{\KetBra{+}{+}+\KetBra{-}{-},\rho\right\}\right]
\end{eqnarray}
\end{widetext}

\begin{figure}
\includegraphics[width=0.30\textwidth, angle=0]{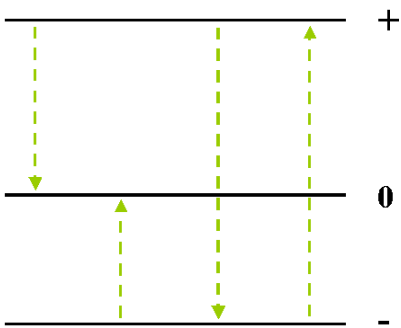}
\caption{(Color online). Scheme of the decays for the dressed
states. There are transitions from $\Ket{+}$ to $\Ket{-}$ and vice
versa. Both $\Ket{+}$ and $\Ket{-}$ decay toward $\Ket{0}$. The
dephasing between the states $\Ket{+}$ and $\Ket{-}$ is not
represented.} \label{fig:dressed_decay_scheme}
\end{figure}

This equation shows that at zero temperature there are the following processes: transitions from $\Ket{+}$ to $\Ket{-}$ and
vice versa, transitions from $\Ket{+}$ to $\Ket{0}$ and from
$\Ket{-}$ to $\Ket{0}$, and a dephasing process involving levels
$\Ket{+}$ and $\Ket{-}$ (see figure \ref{fig:dressed_decay_scheme}). This suggests the idea that the
damping can help to transfer population to level $\Ket{0}$, so
that the efficiency of the counterintuitive sequence should be positively affected by the dissipation.

\section{Analysis of the Efficiency at Zero Temperature}

In this section we analyze the efficiency of both STIRAP and
b-STIRAP processes, by numerically studying the post-pulse
population of the target state $\Ket{3}$ and compare the
prediction of our model with the predictions of a phenomenological
model introduced in ref. \cite{STIRAP spontaneous emission}.

We consider Gaussian laser pulses:
\begin{subequations}
\begin{eqnarray}
\Omega_1&=&\frac{\Omega_0}{2}\, e^{-(t-\tau/2)^2/T^2}\,,\\
\Omega_2&=&\frac{\Omega_0}{2}\, e^{-(t+\tau/2)^2/T^2}\,,
\end{eqnarray}
\end{subequations}
taking into account that we have the so called intuitive sequence
(which corresponds to b-STIRAP) when $\Omega_p=\Omega_2$ and
$\Omega_s=\Omega_1$, while we get the counterintuitive sequence
(which corresponds to STIRAP) when $\Omega_p=\Omega_1$ and
$\Omega_s=\Omega_2$.

The phenomenological model which we will compare with our
microscopic model corresponds to the following master equation:
\begin{eqnarray}
\dot\rho = -i[H,\rho]-\frac{1}{2} D\,,
\end{eqnarray}
with
\begin{eqnarray}
\nonumber
D=\left( %
\begin{array}{ccc}
-2\Gamma_1\rho_{22} & (\Gamma_1+\Gamma_3)\rho_{12} & 0 \\
(\Gamma_1+\Gamma_3)\rho_{21} & 2(\Gamma_1+\Gamma_3)\rho_{22} & (\Gamma_1+\Gamma_3)\rho_{23} \\
0 & (\Gamma_1+\Gamma_3)\rho_{32} & -2\Gamma_3\rho_{22}
\end{array}%
\right)\,,\\
\end{eqnarray}
which describes spontaneous emission from level $2$ to levels $1$
and $3$ with rates $\Gamma_1$ and $\Gamma_3$, respectively. Such a
master equation is related to the bare states and then turns out
to be time-independent.

Concerning the microscopic model, we assume flat spectrum for both
the transitions $2\rightarrow 1$, corresponding to $J_a(\omega)$,
and $2\rightarrow 3$, corresponding to $J_b(\omega)$. In
particular we will assume $J_a(\omega)\equiv \Gamma$ and
$J_b(\omega)\equiv \alpha\Gamma$. This may come, for instance,
from the assumption that the dipole moments between the states
$\Ket{1}$ and $\Ket{2}$ and between the states $\Ket{3}$ and
$\Ket{2}$ are proportional, so that $g_k^{(12)}=\alpha g_k^{(32)}$
for any $k$ in eq. (\ref{Hinteraction}).

\subsection{The counterintuitive sequence}

We first analyze the counterintuitive sequence, where the
population is carried by the dark state $\Ket{0}$.

\begin{figure}
\includegraphics[width=0.40\textwidth, angle=0]{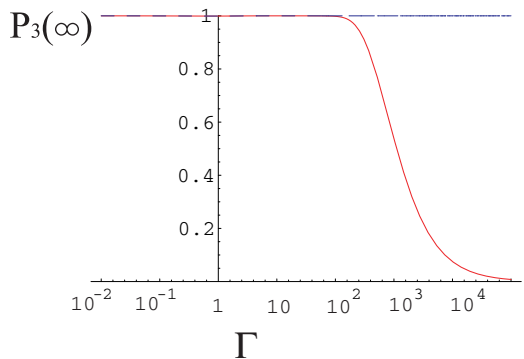}
\caption{(Color online). Counterintuitive sequence. Final
population vs $\Gamma$ (in units of $T^{-1}$ and in logarithmic
scale) according to microscopic (dashed blue line) and
phenomenological (solid red line) model. The relevant parameters
are $\Omega_0=25\, T^{-1}$, $\tau=1.5\, T^{-1}$, $T\,\Delta=1$,
$\alpha=1$.} \label{fig:counterintuitive_compare}
\end{figure}

Figure \ref{fig:counterintuitive_compare} shows the comparison
between the microscopic and the phenomenological models, with
$\alpha=1$ and $\Gamma_1=\Gamma_2=\Gamma$. It is evident that the
microscopic model predicts a very high efficiency (essentially
one) for a wider range of $\Gamma$. This can be explained on the
basis of the decay scheme in fig. \ref{fig:dressed_decay_scheme}:
all the decay processes describe either jumps toward $\Ket{0}$ or
toward states which in turn decay toward $\Ket{0}$, so that the
dark state is robust against zero temperature dissipation.

\begin{figure}
\includegraphics[width=0.40\textwidth, angle=0]{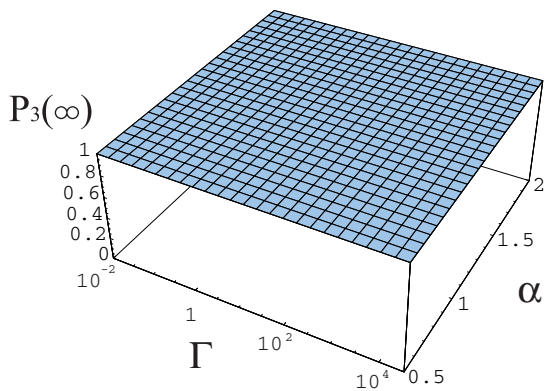}
\caption{(Color online). Counterintuitive sequence. Final
population vs both $\Gamma$ (in units of $T^{-1}$ and in
logarithmic scale) and $\alpha$ according to the microscopic
model. The relevant parameters are $\Omega_0= 25\, T^{-1}$,
$\tau=1.5\,T^{-1}$, $T\,\Delta=1$.}
\label{fig:counterintuitive_different_dipoles}
\end{figure}

The robustness of the counterintuitive scheme is not related to
the special choice $\alpha=1$. Indeed, figure
\ref{fig:counterintuitive_different_dipoles} shows the dependence
of the post-pulse population of state $\Ket{3}$ on both $\Gamma$
and the parameter $\alpha$ which characterizes the difference in
the intensities of the dipolar coupling constant involving
different couples of levels. It is quite evident that the
efficiency of the scheme is not affected by a discrepancy in the
decay rates.

\subsection{The intuitive sequence}

\begin{figure}
\includegraphics[width=0.40\textwidth, angle=0]{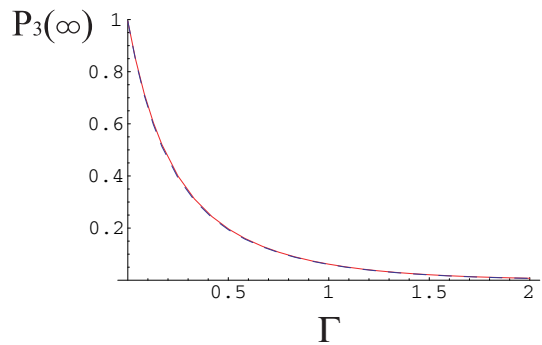}
\caption{(Color online). Intuitive sequence. Final population vs
$\Gamma$ (in units of $T^{-1}$) according to microscopic (dashed
blue line) and phenomenological (solid red line) model. The
relevant parameters are $\Omega_0=25\, T^{-1}$, $\tau=1.5\,
T^{-1}$, $T\,\Delta=1$, $\alpha=1$. The two curves are essentially
coincident.} \label{fig:intuitive_compare}
\end{figure}

Concerning the b-STIRAP process (i.e. in the intuitive sequence),
we find that the two models predict very similar results. In
particular, fig. \ref{fig:intuitive_compare} shows the perfect
coincidence of the predictions of the two models in the case
$\alpha=1$ and $\Gamma_1=\Gamma_2=\Gamma$. Moreover, from this
figure one can see that (for both models) the efficiency is very
sensitive to the presence of decays, so that it almost drops to
zero at $\Gamma\,T=1$. The reason of the fragility of the
efficiency in this scheme is that, while all the populations are
guided by the decay towards the state $\Ket{0}$, population
transfer is instead carried on by the state $\Ket{-}$.

It is worth noting that the result of the comparison is
qualitatively quite similar to the result of the comparison we
made in connection to the scheme with external decay
\cite{ref:Scala2010}. Indeed, in both cases the predictions from
the microscopic and the phenomenological models are almost
coincident for the intuitive sequence, while for the
counterintuitive sequence we find that the microscopic model
predicts a higher efficiency. Nevertheless, we stress here the
fact that the enhancement of efficiency in STIRAP in the strong
damping limit is due to very different mechanisms in the two cases
of external and internal decay. In fact, while for external decay
a very strong damping is responsible for a dynamical decoupling of
the dark state, which then is protected against losses, in the
case of internal decay the dissipation is instead responsible for
transitions toward the state that carries the population,
therefore protecting the process of population transfer.

\section{Analysis of the Efficiency at Nonzero Temperature}

In this section we consider the effects of nonzero temperature.
Looking at \eqref{eq:EnneDiOmega}, we see that the $N(\omega)$'s
are evaluated at very close frequencies, which are essentially
$\omega\approx\omega_{21}\approx\omega_{23}$. For this reason, we
described temperature by a single number $N$ which is the number
of photons in the reservoir modes of frequencies close to the bare
atomic transitions.

We have seen that at zero temperature, for increasing $\Gamma$
(and even for quite small values of the decay constant) the
efficiency falls down to zero when the intuitive sequence is
applied. When temperature is non-vanishing all the transitions
included in \eqref{eq:MasterEq_NonZeroTemperature} but not present
in \eqref{eq:MasterEq_ZeroTemperature} must be considered. In
particular, transitions from $\Ket{0}$ to $\Ket{-}$ should
increase the efficiency of the b-STIRAP process, since in this
scheme the population is transferred through the state $\Ket{-}$.
On the other hand, in the counterintuitive sequence we should get
a lower transfer efficiency, since thermal terms are responsible
for loss of population from state $\Ket{0}$ during the STIRAP
process.

\begin{figure}
\subfigure[]{\includegraphics[width=0.40\textwidth,
angle=0,clip=a]{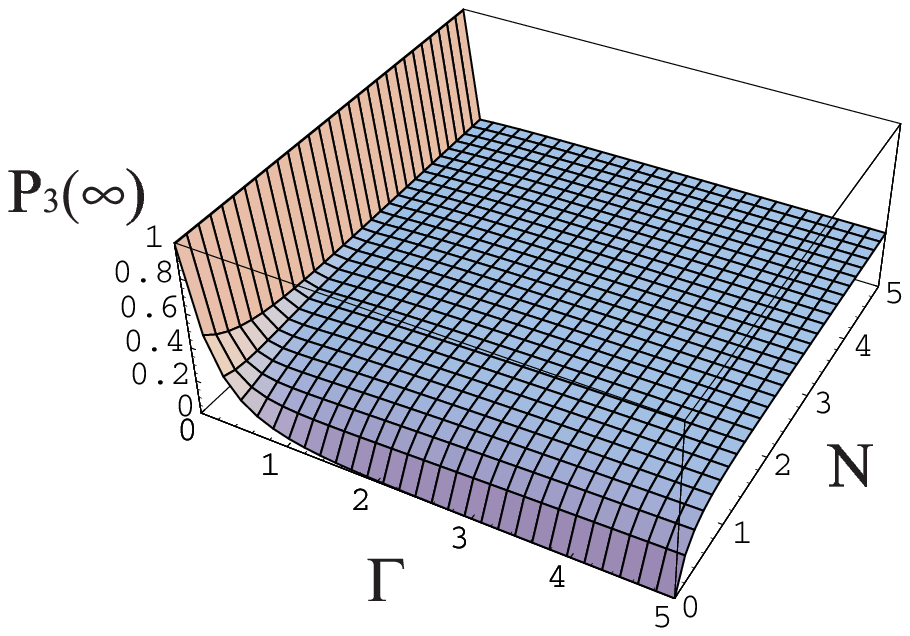}} %
\subfigure[]{\includegraphics[width=0.40\textwidth,
angle=0,clip=b ]{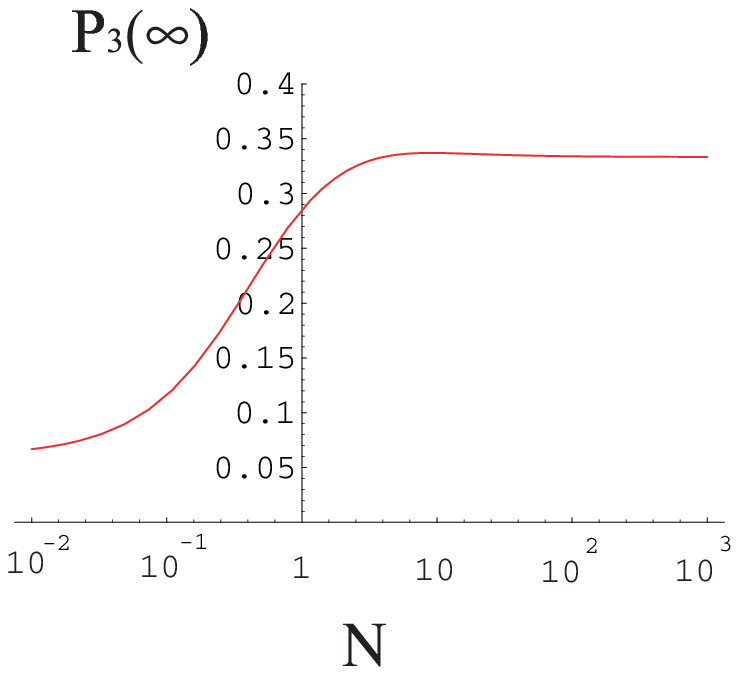}}%
\caption{(Color online). Intuitive sequence. (a) Final population
vs $\Gamma$ (in units of $T^{-1}$) and the number of photons $N$.
(b) Final population vs the number of photons $N$ (in logarithmic
scale) , for $\Gamma=1$. In both cases, the relevant parameters
are $\Omega_0=25\, T^{-1}$, $\tau=1.5\, T^{-1}$, $T\,\Delta=1$,
$\alpha=1$. } \label{fig:intuitive_compare_NonZeroT}
\end{figure}

Figure \ref{fig:intuitive_compare_NonZeroT}a shows the dependence
of the post-pulse population of the state $\Ket{3}$ on $\Gamma$
and temperature (through the number of photons $N$ in the relevant
reservoir modes), for the intuitive sequence. It is well visible
that the efficiency, which goes to zero for large $\Gamma$ in the
zero temperature regime, reaches nonzero values for non vanishing
temperature. Such efficiency reaches a maximum value for
intermediate values of temperature. As an example, figure
\ref{fig:intuitive_compare_NonZeroT}b shows the temperature
dependence (in a wider range) of the efficiency for $\Gamma=1$: in
this case the optimal point is reached at $N\simeq 10$.

\begin{figure}
\includegraphics[width=0.40\textwidth, angle=0]{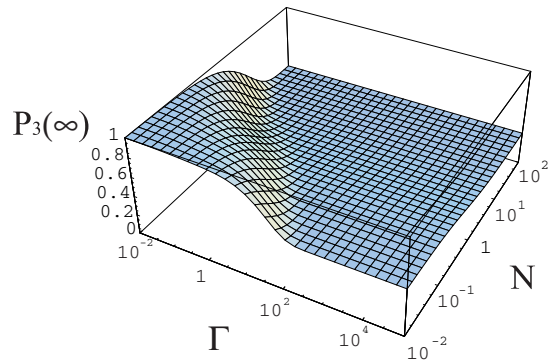}
\caption{(Color online). Counterintuitive sequence. Final
population vs $\Gamma$ (in units of $T^{-1}$ and in logarithmic
scale) and the number of photons $N$ (in logarithmic scale). The
relevant parameters are $\Omega_0=25\, T^{-1}$, $\tau=1.5\,
T^{-1}$, $T\,\Delta=1$, $\alpha=1$. }
\label{fig:counterintuitive_compare_NonZeroT}
\end{figure}

Figure \ref{fig:counterintuitive_compare_NonZeroT} shows the
post-pulse population of the state $\Ket{3}$ on $\Gamma$ and
temperature for the counterintuitive sequence. In this case it is
well visible that the temperature negatively affects the
efficiency of the population transfer. Indeed, even a very small
amount of thermal photons is responsible for a significant
diminishing of the post-pulse population, which instead approaches
unity at rigorously zero temperature.

\section{Discussion and Conclusive Remarks}

In this paper we have analyzed the effects on a STIRAP scheme of
the losses of the auxiliary level towards the two metastable
states, examined by means of the numerical resolution of a master
equation which has been microscopically derived. It is worth
noting that, in order to remove the rapid oscillations in the
system Hamiltonian, we are forced to describe the system in a
rotating frame, where the system-bath interaction term turns out
to be time-dependent. Therefore, when deriving the master equation
we have to deal both with a slowly varying system Hamiltonian,
which can be treated following the general theory by Davies and
Spohn, and a rapidly oscillating system-environment interaction
term. This last point makes the final master equation different
from what one usually gets. Indeed the zero temperature
transitions are not guiding the system towards the dressed
(rotating) ground state $\Ket{-}$, but instead towards the state
$\Ket{0}$. In this sense, the oscillating terms in the
system-environment interaction act like a pumping which makes the
counterintuitive sequence much more robust than the intuitive
sequence.

The inclusion of the nonzero temperature terms in the master
equation partially modifies these conclusions. In fact, thermal
photons switch on different transitions, which make the post-pulse
population of state $\Ket{-}$ different from zero. As a
consequence the efficiency of the counterintuitive sequence is
reduced, while on the other hand the efficiency of the intuitive
sequence increases. However, such an increasing is not high enough
to make the intuitive sequence preferable to the counterintuitive
one.

Though the analysis of this STIRAP scheme (with the same decay
channels), performed in previous papers by means of
phenomenological dissipative terms, has shown a high robustness of
the counterintuitive sequence with respect to losses, our analysis
shows that the scheme at zero temperature is much more robust for
very high decay rates. This is due to the fact that, as already
pointed out, the zero temperature decay channels force the system
to the state $\Ket{0}$, which is the population carrier in this
sequence.

\section*{Acknowledgements}

This work is supported by European Commission's projects EMALI and
FASTQUAST, and the Bulgarian Science Fund grants VU-F-205/06,
VU-I-301/07, and D002-90/08. Support from MIUR Project N.
II04C0E3F3 is also acknowledged.

\appendix

\section{Derivation of the Master Equation}\label{AppDerivation}

In this appendix we give some more details about the derivation of
the master equation in our model. In our case we consider:
\begin{subequations}
\begin{eqnarray}
A_{a}^{+}&=&\KetBra{1}{2}\,,\\
A_{a}^{-}&=&\KetBra{2}{1}\,,\\
A_{b}^{+}&=&\KetBra{3}{2}\,,\\
A_{b}^{-}&=&\KetBra{2}{3}\,,
\end{eqnarray}
\end{subequations}
\begin{subequations}
\begin{eqnarray}
B_{a}=\sum_k g_k^{(12)}(b_k+b_k^\dag)\,,\label{bathoperatorsa}\\
B_{b}=\sum_k g_k^{(32)}(b_k+b_k^\dag)\,\label{bathoperatorsb},
\end{eqnarray}
\end{subequations}
from which we obtain the following jump operators:
\begin{widetext}
\begin{eqnarray}
\nonumber &&
\begin{array}{ll}
A_{a}^{+}(\omega_{+0}) = \cos\theta\cos\varphi\KetBra{0}{+} \,, &
A_{a}^{-}(\omega_{+0}) = 0\,, \\
A_{b}^{+}(\omega_{+0}) = -\sin\theta\cos\varphi\KetBra{0}{+}\,, &
A_{b}^{-}(\omega_{+0}) = 0\,, \\
A_{a}^{+}(\omega_{0-}) = 0\,, &
A_{a}^{-}(\omega_{0-}) = -\cos\theta\sin\varphi\KetBra{-}{0} \,,\\
A_{b}^{+}(\omega_{0-}) = 0\,, &
A_{b}^{-}(\omega_{0-}) = \sin\theta\sin\varphi\KetBra{-}{0}\,,\\
A_{a}^{+}(\omega_{+-}) = \sin\theta\cos^2\varphi\KetBra{-}{+} \,,
&
A_{a}^{-}(\omega_{+-}) = -\sin\theta\sin^2\varphi\KetBra{-}{+}\,, \\
A_{b}^{+}(\omega_{+-}) = \cos\theta\cos^2\varphi\KetBra{-}{+}\,, &
A_{b}^{-}(\omega_{+-}) = -\cos\theta\sin^2\varphi\KetBra{-}{+}\,,\\
\end{array}\\
\nonumber
&&\,\,A_{a}^{+}(0) = A_{a}^{-}(0) = \sin\theta\sin\varphi\cos\varphi\left(\KetBra{+}{+}-\KetBra{-}{-}\right)\,, \\
&&\,\,A_{b}^{+}(0) = A_{b}^{-}(0) =
\cos\theta\sin\varphi\cos\varphi\left(\KetBra{+}{+}-\KetBra{-}{-}\right)
\,.
\end{eqnarray}
\end{widetext}

Putting all the jump operators inside the Lindblad form in
\eqref{eq:ME_General_Lindb}, we get the master equation in
\eqref{eq:MasterEq_NonZeroTemperature}.

\end{document}